\documentclass[onecolumn,aps,prl,showpacs,superscriptaddress,notitlepage,nofootinbib]{revtex4-1}
\usepackage[a4paper,top=3cm,bottom=3cm,left=3.6cm,right=3.6cm]{geometry}
\usepackage[english]{babel}
\usepackage{graphicx}				
\usepackage{booktabs}			
\usepackage{amsmath}				
\usepackage{mathtools}				
\usepackage{braket}					
\usepackage{multirow}
\usepackage{setspace}

\DeclarePairedDelimiter\abs{\lvert}{\rvert}

\makeatletter
\let\cat@comma@active\@empty
\makeatother

\begin{document}
\title{Pure down-conversion photons through sub-coherence length domain engineering}

\author{Francesco Graffitti}
\email{fg13@hw.ac.uk} 
\affiliation{Scottish Universities Physics Alliance (SUPA), Institute of Photonics and Quantum Sciences, School of Engineering and Physical Sciences, Heriot-Watt University, Edinburgh EH14 4AS, UK}
\author{Dmytro Kundys}
\affiliation{Scottish Universities Physics Alliance (SUPA), Institute of Photonics and Quantum Sciences, School of Engineering and Physical Sciences, Heriot-Watt University, Edinburgh EH14 4AS, UK}
\author{Derryck T. Reid}
\affiliation{Scottish Universities Physics Alliance (SUPA), Institute of Photonics and Quantum Sciences, School of Engineering and Physical Sciences, Heriot-Watt University, Edinburgh EH14 4AS, UK}
\author{Agata M. Bra\'nczyk}
\affiliation{Perimeter Institute for Theoretical Physics, Waterloo, Ontario, N2L 2Y5, Canada}
\author{Alessandro Fedrizzi}
\affiliation{Scottish Universities Physics Alliance (SUPA), Institute of Photonics and Quantum Sciences, School of Engineering and Physical Sciences, Heriot-Watt University, Edinburgh EH14 4AS, UK}

\begin{abstract}
 Photonic quantum technology relies on efficient sources of coherent single photons, the ideal carriers of quantum information. Heralded single photons from parametric down-conversion can approximate on-demand single photons to a desired degree, with high spectral purities achieved through group-velocity matching and tailored crystal nonlinearities.
 Here we propose crystal nonlinearity engineering techniques with sub-coherence-length domains. We first introduce a combination of two existing methods: a deterministic approach with coherence-length domains and probabilistic domain-width annealing. We then show how the same deterministic domain-flip approach can be implemented with sub-coherence length domains. Both of these complementary techniques create highly pure photons, outperforming previous methods, in particular for short nonlinear crystals matched to femtosecond lasers.
\end{abstract}

\maketitle

\section{Introduction}
The generation of pure single photons on demand is a key requirement for photonic quantum technologies. Single-photon sources can be approximated using either nonlinear parametric processes or single quantum emitters such as quantum dots \cite{lounis2005single,shields2007semiconductor,Michler2009}.
The most widely used technique for generating single photons is parametric down-conversion (PDC) in nonlinear optical crystals \cite{burnham1970observation,kwiat1995new,tanzilli2001highly,mosley2008heralded}. 
In a PDC process, a pump laser is sent through a crystal with large $\chi^{(2)}$ nonlinearity, where $\chi^{(2)}$ is the second order nonlinear electric susceptibility. A pump photon has a small probability of being annihilated while creating pairs of ``down-converted'' single photons under conservation of energy and momentum. Due to the spontaneous nature of photon-pair creation, the complete generated state is described as a superposition of a large vacuum component, the desired photon pairs, as well as typically undesired higher-order photon-pair events \cite{jennewein2011single, christ2012limits}. 

PDC can be used as a heralded source of single photons, in which one photon of a pair is sacrificed in detection to herald the presence of a photon in the other mode.
The heralded photon generally emerges in a spectrally mixed state due to strong (anti)correlation in the joint spectrum of PDC photon pairs resulting from the conservation of energy and momentum.
To create pure photons, we therefore need to eliminate correlations without introducing mixture in other degrees of freedom.

The easiest way to reduce spectral correlations is to employ narrowband spectral filters. This approach, however, severely compromises the heralding efficiency  (i.e. the probability of detecting a photon knowing that the other photon of the pair has been detected)  of the source even with ideal filters~\cite{meyer2017filtering} and decreases the absolute flux of the heralded photons.   At high pump powers, narrowband filtering can also introduce mixing in the photon-number degree of freedom\cite{branczyk2010optimized}.

A lossless method to remove frequency correlations is to use group-velocity matching (GVM) \cite{keller1997theory,grice2001eliminating,mosley2008heralded,u2006generation}. Starting with a set of desired pump and down-conversion wavelengths, one can find phasematching conditions in certain crystals that allow the inverse group velocity of the pump laser in the nonlinear crystal to either match one of the PDC photon's inverse group velocities or to match the average of the two PDC photons' inverse group velocities. This erases the timing information that otherwise arises between the pump and PDC photons which in consequence leads to more separable joint spectra and high purity photons.
GVM can be achieved both in bulk crystals and in periodically poled crystals, where the crystal's axes are flipped every coherence length (i.e. the distance over which the phase between the pump and the PDC photons changes by $\pi$) to fulfil the quasi-phase matching condition \cite{fejer1992quasi}, resulting in a periodical configuration of domains with alternating orientation.

What group-velocity matching cannot address is that PDC in standard nonlinear crystals with a rectangular nonlinearity profile produces \emph{sinc}-shaped joint spectral amplitudes (JSA) which further restricts photon purity \cite{branczyk2010optimized}.
The first step towards overcoming this issue was proposed by Bra\'nczyk \emph{et al.}, who showed that the nonlinearity profile of the PDC crystal could be suitably shaped via domain engineering~\cite{branczyk2011engineered} in order to achieve a Gaussian phase matching function (PMF).
Considering a poled crystal, it is indeed possible to individually choose each domain orientation such that the effective PMF fits an almost arbitrary target function. 
A proof-of-principle experiment~\cite{branczyk2011engineered} showed a high overlap between the design and the experimentally determined PMF, proving the feasibility of tailored nonlinearities for this purpose.

This technique was subsequently refined \cite{dixon2013spectral,dosseva2016shaping,tambasco2016domain}, leading to even better approximations to ideal nonlinearity profiles. Simulations showed that these methods would increase photon purities without compromising the heralding efficiency of the source.
Chen \emph{et al.} \cite{chen2017efficient} showed that the spectral response of a tailored PDC crystal had a low amount of entanglement in the frequency degree of freedom. 
However, a direct measurement of the indistinguishability (such as Hong-Ou-Mandel (HOM) interference) of photons produced by two independent PDC processes in engineered crystals has not been done yet. Note that HOM interference between photons generated from a single pulse in one crystal is not a good indicator for single photon purity because one can achieve perfect interference also with a standard periodically poled crystal \cite{hong1987measurement}.

All domain engineering methods developed so far are intrinsically related to a coherence-length structure typical of periodically poled crystals.
In \cite{dixon2013spectral} the domain-periodicity is restricted to twice the coherence length of the crystal; in the other methods \cite{branczyk2011engineered,dosseva2016shaping,tambasco2016domain}, the domain-width is kept fixed to the coherence length.
While this is an intuitive choice and allows a simple approach to  PMF shaping, the fixed structure substantially limits  the performance and versatility of domain engineering techniques. 
In this work we move beyond the coherence-length domain boundaries to allow for more fine-grained shaping of the nonlinearity. To this aim we explore two complementary approaches.

In the first approach, we modify a deterministic method for domain engineering \cite{tambasco2016domain} and  use it for tailoring a crystal with fixed domain widths: we then shift the boundaries of the domains by means of a previously existing annealing method developed for classical applications in higher-harmonic generation \cite{reid2003engineered,kornaszewski2008designer}. In the second approach, we generalise the algorithm in \cite{tambasco2016domain} to arbitrarily small, but constant, domain widths (not necessarily equal to the coherence length) and to a complex field amplitude target function.

Both  methods lead to a better approximation of the desired phase matching function and to an enhanced heralded single-photon purity, especially in the short-crystal regime.

\section{Joint spectral amplitude}
The output of a second-order nonlinear crystal, pumped by a bright coherent field with envelope function $\alpha(\omega_s+\omega_i)$, can be written as \cite{harder2016optimized}:
\begin{align}\label{eq:fullpdc}
\ket{\psi}_{s,i}=\mathcal{T}\exp\left(2\pi B O \iint d\omega_s d\omega_i  \phi(\Delta k(\omega_s,\omega_i ))  \alpha(\omega_s+\omega_i)\hat{a}_s^{\dagger}(\omega_s)\hat{a}_i^{\dagger}(\omega_i) -h.c.\right)\ket{0}_{s,i}\,,
\end{align}
where  ``$p$'' labels the pump field and ``$s$'' and ``$i$'' label the two down-converted ``signal'' and ``idler'' photons respectively, and $\mathcal{T}$ is the time-ordering operator \cite{branczyk2011time,christ2013theory,quesada2014effects,quesada2015time}. We call $\phi(\Delta k(\omega_s,\omega_i ))$ the phase matching function and define it below.
 The constant $O$ is related to the transverse modes of the photons \cite{harder2016optimized}, and ${B=\chi^{(2)}_{0}(\hbar\omega_{0p}\omega_{0s}\omega_{0i}/\epsilon_0\pi^{3}c^{3}n_pn_sn_i)^{1/2}}/2$,
 where $\omega_{0,j}$ (with $j=p,i,s$) are the central frequencies of the fields, $n_j$ are the corresponding refractive indices and $\chi^{(2)}_{0}$ is the crystal's second order nonlinearity.
 There exist a number of related, but slightly different, definitions of the phase matching function in the literature. We define it as:
\begin{equation}
\phi(\Delta k(\omega_s,\omega_i )) = \int_{-\infty}^{+\infty} g(z) \ e^{i \Delta k (\omega_s,\omega_i ) z} dz\,,
\label{generalPMF}
\end{equation}
where $\Delta k (\omega_s,\omega_i)= k_p  (\omega_s+ \omega_i) - k_s (\omega_s)- k_i (\omega_i)$ is the phase mismatch, which depends on the material dispersion, and $g(z) = \chi^{(2)}(z)/\chi^{(2)}_0$ is the normalised nonlinearity along the crystal which will be effectively ``tailored'' via domain engineering. We note that $g(z)$ is a real function. When defined as above, the PMF for two consecutive domains is the sum of the PMFs for the individual domains\footnote{Take for example a block of two domains. If the normalized nonlinearity function for the $i$th domain ($i=1,2$) is $g_i(z)$, then the normalized nonlinearity function for both domains is $g_{1,2}(z)=g_1(z)+g_2(z)$. According to our definition, the PMF for both domains is then  $\phi_{12}(\Delta k)= \int_{-\infty}^{+\infty} g_{12}(z) \ e^{i \Delta k z} dz= \int_{-\infty}^{+\infty} g_{1}(z) \ e^{i \Delta k z} dz+ \int_{-\infty}^{+\infty} g_{2}(z) \ e^{i \Delta k z} dz=\phi_{1}(\Delta k)+\phi_{2}(\Delta k)$, where $\phi_{i}(\Delta k)$ is the PMF for the $i$th domain. Take on the other hand another popular definition for the PMF, $\phi^{\text{other}}(\Delta k) =L^{-1} \int_{-\infty}^{+\infty} g(z) \ e^{i \Delta k  z} dz$. The function $\phi^{\text{other}}(\Delta k) $ is dimensionless, but in this case $\phi^{\text{other}}_{12}(\Delta k)=(2l_c)^{-1} \int_{-\infty}^{+\infty} g_{12}(z) \ e^{i \Delta k z} dz \neq\phi^{\text{other}}_{1}(\Delta k)+\phi^{\text{other}}_{2}(\Delta k)= l_c^{-1} \int_{-\infty}^{+\infty} g_{1}(z) \ e^{i \Delta k z} dz+l_c^{-1}  \int_{-\infty}^{+\infty} g_{2}(z) \ e^{i \Delta k z} $.} (this definition has units of length, while in other popular definitions, the PMF is dimensionless).  

The joint spectrum of the down-converted light is given by the product of the PMF and the pump envelope function. We therefore define the \emph{joint spectral amplitude} as ${f(\omega_s,\omega_i )\equiv\phi(\Delta k(\omega_s,\omega_i ))  \alpha(\omega_s+\omega_i)}$. 
For the purposes of heralding single photons, we can restrict our analysis to the two-photon state
\begin{equation}
\ket{\psi_{\text{pair}}}_{s,i}= 2\pi BO \iint d\omega_s d\omega_i \ f(\omega_s,\omega_i ) \hat{a}_s^{\dagger}(\omega_s)\hat{a}_i^{\dagger}(\omega_i) \ket{0}_{s,i}\,, \label{mainstate}
\end{equation}
given by the first-order term in the Dyson series expansion of Eq. (\ref{eq:fullpdc}). 

The spectral purity of the heralded single-photon state is defined as $P=\sum_k b_k^4$ where $b_k$ are the Schmidt coefficients defined by the Schmidt decomposition $f(\omega_s,\omega_i ) = \sum_k b_k q_k (\omega_s) r_k(\omega_i)$. The purity is therefore limited by correlations in the joint spectral amplitude.
 We thus seek a separable joint spectral amplitude, that is, a function of the form  $f(\omega_s,\omega_i ) = q(\omega_s) r(\omega_i )$. For a symmetric joint spectral amplitude, the standard approach is to first choose group velocities that make $\phi(\Delta k(\omega_s,\omega_i ))$ intersect $\alpha(\omega_s+\omega_i)$ at right angles (when plotted as a function of $\omega_i$ and $\omega_s$), then match the widths of the pump envelope and phase matching functions \cite{laudenbach2016modelling}. This removes most of the correlations, but those that are due to the shape of the PMF remain. Spectral filtering can be used to remove the final correlations, but very high spectral purities can only be achieved at the expense of the heralding efficiency. To achieve high purities \emph{and} high heralding efficiencies, the phase matching function must be suitably engineered.

\section{Engineering the phase matching function}

In a periodically-poled crystal,  $g(z)$ alternates between $+1$ and $-1$ every coherence length: for a large number of periods $g(z)$ can be approximated as constant and the PMF is consequently proportional to the \emph{sinc} function.

Aperiodic poling is also possible.  Although such techniques still constrain $g(z)$ to values of $+1$ and $-1$, the non-trivial structure makes it possible to shape the effective nonlinearity of the crystal and consequently to customise the PMF. In particular, non-trivial poling can make the PMF approximate a Gaussian function, which is necessary to make $f(\omega_s,\omega_i )$ separable.

Domain engineering methods have long been studied in nonlinear optics, for example to compress or shape pulses in second harmonic generation \cite{arbore1997engineerable,imeshev2000ultrashort,shur2001recent,zhang2001optimal,reid2003engineered,kornaszewski2008designer}.
These techniques have only recently been adapted for single-photon generation. In this context, existing methods for non-trivial poling fall into two categories. Those that vary the domain widths of a predefined poling configuration \cite{dixon2013spectral}, and those that keep the domain widths equal, but vary their relative orientations \cite{branczyk2011engineered,dosseva2016shaping,tambasco2016domain}. All methods in the latter category have thus far used domains equal in width to the crystal's coherence length, where the coherence length $\ell_c$ is defined in terms of the phase mismatch $\Delta k (\omega_{s,0}, \omega_{i,0})$ at the central PDC frequencies as follows: $\ell_c = \pi / \Delta k(\omega_{s,0},\omega_{i,0})$.

In the following, we describe two methods for engineering $g(z)$. Both use the method recently introduced by Tambasco  \cite{tambasco2016domain} as a starting point, but deviate from the method by allowing domain widths smaller than the crystal's coherence length.  This move toward sub-coherence length structures allows much greater accuracy in tailoring the phase matching function.

\subsection{Deterministic domain engineering with fixed domain widths}
We begin by summarising the method introduced by Tambasco \emph{et al.} \cite{tambasco2016domain}. In this method the width of each domain in the grating is fixed, and the nonlinearity profile is shaped by choosing the orientation of successive ``poling blocks'' along the crystal, starting with the crystal's input face (the term ``poling block'' refers to two consecutive domains of fixed length $w=\ell_c$, where $\ell_c$ is the coherence length and $w$ is the domain width). The decision to flip (or not flip) a given block is determined by which option gives a closer approximation to the target signal and idler joint field at that point in the crystal. The normalised field inside the crystal at position $z$ is defined as 
\begin{equation}
A(z,\Delta k) = -i \int_0^z  g(z')\  e^{i \Delta k z'} dz' \ ,
\label{amplitude}
\end{equation}
where we have omitted the amplitude prefactor \cite{boyd2003nonlinear}.

The goal is to design a crystal with a $g(z)$ that gives the desired PMF according to Eq. (\ref{generalPMF}). We therefore define a  $g_{\text{target}}(z)$, which in turn defines a target field amplitude 
\begin{equation}
A_{\text{target}} (z,\Delta k) = -i \int_0^z  g_{\text{target}}(z')\  e^{i \Delta k z'} dz' \ .
\label{amplitude}
\end{equation}

It is sufficient to match the  the amplitude at a single value of $\Delta k$.  At $\Delta k=\pi/\ell_c$, the propagation of the pump for a distance equal to one domain  width changes the \emph{real} part of the generated field amplitude by a factor $\Delta A = \pm 2w/\pi$, where the sign depends on the overall previous domain configuration in the crystal. The \emph{imaginary} part of the field is always equal to zero at domain boundaries.  The  poling pattern can be chosen so that the generated field amplitude $A (z,\pi/\ell_c) $ along the crystal fits as well as possible the target field amplitude $A_{\text{target}} (z,\pi/\ell_c) $. For simplicity of notation we will refer to a domain pointing up by ``\textsc{up}'' and to a domain down by ``\textsc{down}''.

\begin{figure*}[htb]\center
\includegraphics[width=1\textwidth]{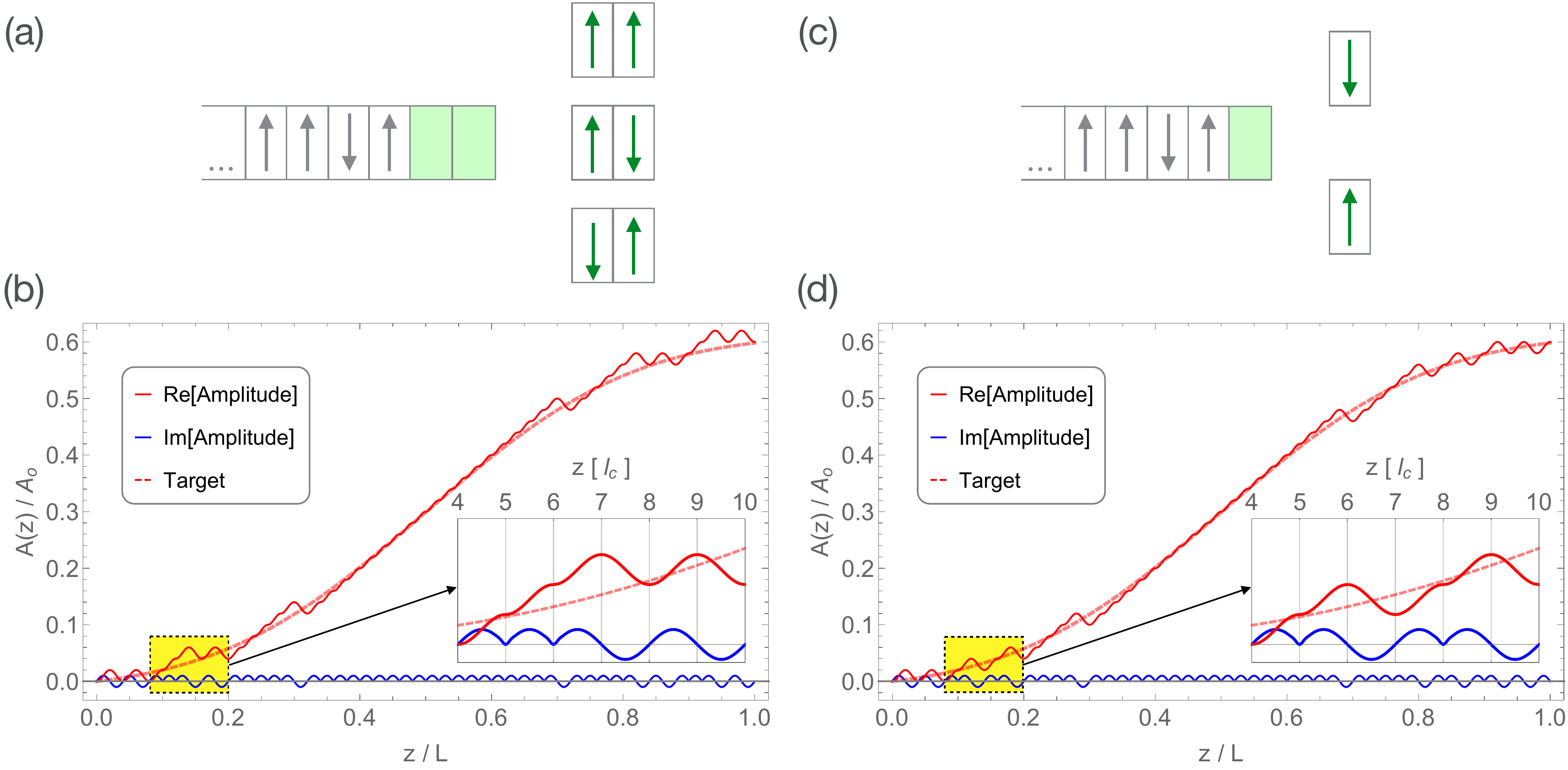}
\caption{
between the algorithm in \cite{tambasco2016domain} executed with blocks of two domains (a,b) and domain by domain (c,d). 
(a,c) Possible configurations in the two methods: in green (grey) are represented the "active" ("inactive") elements, i.e. the elements which are (are not) influenced by the current step of the algorithm. In the first case there are three possible configurations which are independent from the previous two-domains orientation. In the second case at each step a domain \textsc{up} or a domain \textsc{down} is added according to the four conditions described in the main text.
(b,d) The two rules applied to a crystal composed of 50 domains. The dashed red line is the target function and the oscillatory functions are the real and imaginary part of the field amplitude along the crystal, normalised with respect to the amplitude $A_0$ at the end of a standard periodically poled crystal having the same spectral width. 
The highlighted sections show how choosing the domain orientation one by one instead of using block of two domains allows to fit the target function more accurately, and that the imaginary part of the amplitude cancels out every domain length.}
\label{TambascoVSimproved}
\end{figure*}

The poling algorithm proposed by Tambasco \emph{et al}. involves three different building blocks which are chosen to track the target field amplitude: an \textsc{up}-\textsc{up} block leaves the average field amplitude unchanged, an \textsc{up}-\textsc{down} block increases the average field amplitude by a factor $2 \Delta A$, while \textsc{down}-\textsc{up} decrease the average field amplitude by $2 \Delta A$. These three possible configurations are graphically represented in Fig. \ref{TambascoVSimproved} (a).

Tambasco \emph{et al.} framed their algorithm in terms of blocks to ensure that the inverted regions were of equal width: this is not a necessary requirement---one might just as well choose each individual domain's orientation.
We therefore consider the ``domain by domain'' error function
\begin{equation}
e(x+w) = A_{\text{target}}(z+w,\pi/\ell_c)- A(z, \pi/\ell_c)\,,
\end{equation}
which quantifies the difference between the generated field amplitude at a certain position $z$ and the target field amplitude at $z+w$ (i.e. after one domain).
Whether to use a domain \textsc{up} or a domain \textsc{down} is determined by the following conditions:
\begin{itemize}
  \setlength\itemsep{-1em}
\item if $e(z+w) \geq 0$ and $A(z,\pi/\ell_c) \geq A(z-w,\pi/\ell_c)$ (i.e. in the previous domain the field amplitude was increasing), flip the domain orientation with respect to the previous domain: in this way the amplitude field will continue to increase; \\
\item if $e(z+w) \geq 0$ and $A(z,\pi/\ell_c) \leq A(z-w,\pi/\ell_c)$ (i.e. in the previous domain the field amplitude was decreasing), keep the same orientation of the previous domain: in this way the amplitude field will start increasing;\\
\item if $e(z+w) < 0$ and $A(z,\pi/\ell_c) \geq A(z-w,\pi/\ell_c)$, keep the same orientation of the previous domain: in this way the amplitude field will start decreasing; \\
\item if $e(z+w) < 0$ and $A(z,\pi/\ell_c) \leq A(z-w,\pi/\ell_c)$, flip the domain orientation with respect to the previous domain: in this way the amplitude field will continue to decrease. 
\end{itemize}
With this technique, it is possible to generate a wide range of PMF shapes, as long as the corresponding field amplitude does not vary too quickly \cite{tambasco2016domain}.  The computational runtime of this algorithm is minimal (generally of the order of a few seconds).

As discussed above, the case of a Gaussian modulation proportional to $\allowbreak g(z)=\exp(-(z-L/2)^2/2\sigma^2)\cos((\pi/\ell_c) z)$, where $L$ is the length of the crystal, is the most interesting for engineering a pure heralded photon source. Ignoring the quickly oscillating terms, the resulting target field amplitude is

\begin{equation}
A_{\text{target}}\left(z,\pi/\ell_c\right) = -i\ c \ \left( \text{erf}\left( \frac{L-2z}{2\sqrt{2} \ \sigma} \right) - \text{erf} \left(\frac{L}{2 \sqrt{2}\ \sigma}\right) \right)\, ,
\label{gaussianampli}
\end{equation}
where $L$ is the length of the crystal, $\sigma$ is the width of the Gaussian nonlinearity profile and $c$ is a scale factor.  As suggested in \cite{tambasco2016domain}, a value of $ \sigma \approx L/4$ allows to reach high single-photon purity without significant degradation in the source brightness.
Choosing the prefactor $c=\sqrt{2/\pi}\sigma $ guarantees the optimal PMF: a higher value would compromise the purity because the target amplitude's gradient is too high to be efficiently traced by the actual field while a lower value would reduce the effective nonlinearity of the crystal.

Notice that the target field amplitude in Eq. (\ref{gaussianampli}) is purely imaginary, while the method described above allows only the real part of the field amplitude to be customised. In principle, it is therefore impossible to use this technique to approximate Eq. (\ref{gaussianampli}), but in practice, we can approximate Eq. (\ref{gaussianampli}) up to a phase factor of $i$. This results in a $\pi/2$ phase shift in the exponent of Eq. (\ref{eq:fullpdc}), but when considering just the two-photon term, it can be regarded as an irrelevant global phase. 

Fig.   \ref{TambascoVSimproved} shows a comparison between the algorithm in \cite{tambasco2016domain} and our modified algorithm described above. The modified algorithm yields a closer approximation  to $A_{\text{target}}$. Consequently, the PMF will be closer to the target and, in the case of the Gaussian shaping, it will lead to higher purities.

\subsection{Domain width annealing}
The previous methods consider the case of a poled crystal with constant domain width, which allows to easily approach the problem. However, there is no evidence to suggest that a fixed-domain structure leads to an optimal result, and it is reasonable to ask if it is possible to improve the PMF shape by slightly varying the width of each domain.
To this aim we use an adapted version of the simulated annealing algorithm introduced by Reid \emph{et al.} in \cite{reid2003engineered,kornaszewski2008designer}. Annealing algorithms are commonly exploited for finding a global minimum of a given function dependent on multiple parameters \cite{annealing1983gelatt,kirkpatrick1984optimization} by slightly perturbing the system from a suitable starting point, calculating the relative cost function (commonly called energy in analogy with the internal energy in a physical annealing process) and accepting the change with some probability---the higher the energy the lower is the probability of accepting the new configuration. These kinds of algorithms are probabilistic and may require several runs to get an optimal result.

Given a target PMF $\phi_{\text{target}}(\Delta k)$, we first find the initial configuration of domain orientations by means of the method described above.
Adjacent domains with the same orientation are grouped together into bigger blocks. 
Secondly, we define a starting temperature $T$ and a temperature step $\Delta T$ for the annealing algorithm: we found that a value of $T \epsilon \{0.1;10\}$ and $\Delta T = T/100000$ lead to slightly better results.
The width of each block is then perturbed by up to 1\% and the relative PMF is computed.
The perturbation value 1\% is empirically determined: too small values would lead to slow convergence of the algorithm while if the perturbation is too big the algorithm is unstable and doesn't converge to a correct result. It is then possible to find the system energy defined as
\begin{equation}
E = \left( \sum_{\Delta k} \left[ \abs{\phi_{\text{target}}\left( \Delta k \right)} - \abs{\phi \left( \Delta k \right)} \right]^2 \right)^{1/2}\  \text{.}
\end{equation}
If this energy is smaller than the minimum energy recorded so far the new domain widths are recorded as the best configuration, and they are accepted with a probability of $\exp{\left( -E/T \right)}$ (even if they are not the best configuration).
Finally, the temperature is decreased by $\Delta T$ and the algorithm is repeated until the energy becomes smaller than a chosen threshold or the temperature reaches $0$. 
A more detailed description of the algorithm is provided in Appendix A. This algorithm is computationally demanding and may require a few hours for converging to a solution for a crystal having a large number of domains.

One may think to directly use the annealing procedure without using a pre-processed initial configuration as a seed:
however, our annealing algorithm doesn't allow the flipping of domain orientations, and thus a near-optimal result cannot be obtained without a suitable starting configuration. Moreover, it's reasonable to think that the method discussed above provides a good initial domain-orientation because we know that the generated amplitude has to fit the ideal case of Eq. \eqref{gaussianampli}.

\subsection{Domain engineering with arbitrary small sub-coherence length domains}
Until now we have considered domain-widths equal to the coherence length with small variations of this configuration.
In this section we will see that pushing the algorithm beyond this constraint by allowing a finer discretisation of the domain-structure leads to an even better approximation of the target function.

According to Eq. \eqref{amplitude}, we consider the field amplitude at the end of $m$-th domain:
\begin{align}
A_{m}\left(\{s_n\}_{n=1}^{m},\pi/\ell_c\right)={}& -i \sum_{n=1}^ms_n\int_{(n-1)w}^{n w}  e^{i \frac{\pi}{\ell_c} z}dz \nonumber \\
={}&\frac{\ell_c}{\pi}(e^{-i\frac{\pi}{\ell_c}w}-1)\sum_{n=1}^{m}s_ne^{i\frac{\pi}{\ell_c}nw}\,,
\end{align}
where $w$ is the domain width and $s_n= \pm 1$ is the orientation of the $n$-th domain. Note that $A_{m}\left(\{s_n\}_{n=1}^{m},\pi/\ell_c\right)$ depends on the orientations of all domains that come before it (but not those that come after).

For domain-widths equal to the coherence length, the imaginary part of the field is always zero at the domain boundaries, and therefore it always oscillates about zero (see Fig. \ref{TambascoVSimproved}). If the domain-widths differ from the coherence length, however, the phase might get flipped at a place where the imaginary part is non-zero, providing control over both the real and imaginary parts of the amplitude. With this modification, it is now possible to approximate complex field amplitudes. In particular, the field in Eq. (\ref{gaussianampli}) can be approximated up to the correct phase. But to maintain consistency, we will continue to work with the real target function used in the previous section.

To account for the complex nature of the field amplitude, we  define a cost function that we want to minimize at each domain:
\begin{equation}
e_m(\{s_n\}_{n=1}^{m})=\abs{A_{\mathrm{target}}(mw)-A_{m}(\{s_n\}_{n=1}^{m})}\,.
\label{cost}
\end{equation}

Since the target function has a zero imaginary part, the algorithm with the new cost function will force the imaginary component close to zero.

The algorithm can be summarised as follows. First define a domain width $w$, the coherence length $\ell_c$, and the number of domains $N$. Then, starting from a crystal having just one domain, compute the cost function $e_{\textsc{up}}$ for the case where a domain \textsc{up} is added and the cost function $e_{\textsc{down}}$ when a domain \textsc{down} is added. Next, compare the two cost functions: if $e_{\textsc{up}}<e_{\textsc{down}}$ keep the configuration where the \textsc{up} domain was added, otherwise keep the configuration where the \textsc{down} domain was added. Repeat for each subsequent domain.  A more detailed description of the algorithm is provided in Appendix A. The running time of this algorithm can vary from a several seconds to several minutes depending on the length of the crystal and on the domain-widths.

\subsection{Comparison of the algorithms}
In this section we benchmark our domain engineering techniques in a realistic scenario.
We consider a short potassium titanyl phosphate (KTP) crystal because it can be matched with femtosecond lasers \cite{laudenbach2016modelling}, and its optimal GVM condition (pump wavelength of  791 nm and signal/idler wavelengths of 1582 nm) produces photons compatible with telecom technologies.

\begin{figure*}[htb]\center
\includegraphics[width=1\textwidth]{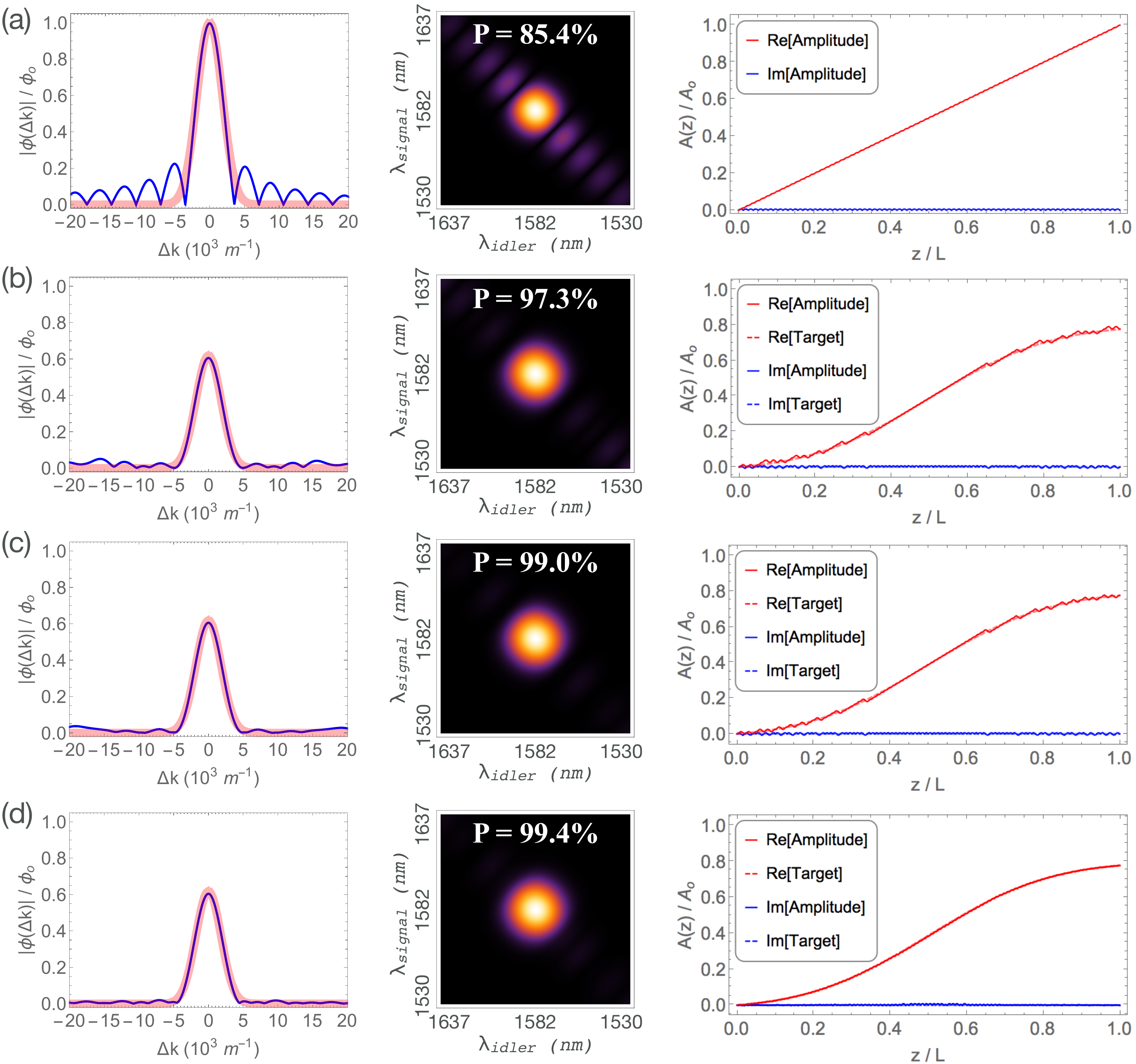}
\caption{
Phase matching functions (first column) in KTP crystals and corresponding joint spectrum and field amplitude along the crystal (second and third columns) in type-II PDC for a short crystal pumped with a Gaussian pulse centred at 791 nm. 
The joint spectrum and the amplitude along are normalised with respect to $\phi_0$ and $A_0$, which are the maximum of the PMF and the maximum amplitude of the periodically poled case, respectively.
(a) Standard ppKTP crystal having a \emph{sinc}-shaped PMF. 
(b) Crystal poling generated using Tambasco fixed domain width algorithm. 
(c) Crystal poling generated by domain width annealing
(d) Crystal poling generated using arbitrary small sub-coherence domains. The domain width equal to a tenth of the coherence length. 
This method achieves single photon purities for short crystals ($99.4\%$) that is very close to the purity of long crystals ($99.5\%$).
 The Sellmeier equations and the temperature-dependent dispersion relation used in the simulations are given in \cite{konig2004extended,fradkin1999tunable,emanueli2003temperature}.
}
\label{comparison}
\end{figure*}

We tailor a short KTP crystal ($\sim 2$ mm) with a Gaussian spectral response using three different domain engineering algorithms: the method by Tambasco \emph{et al.} \cite{tambasco2016domain} and our two new methods. The comparison results are  presented in Fig. \ref{comparison}.
In this regime, the single photon purity increases from a value of $85.4\%$ for a standard periodically poled crystal to $97.3\%$ for the Tambasco algorithm, while a $99.0\%$ photon purity is observed for the width annealing and a value of $99.4\%$ is achieved for arbitrary small sub-coherence domains. 
A practical limitation for sub-coherence length structures is that poling domains cannot be smaller than a couple of microns. The precision to which such domains can be grown is however very high. These practical aspects and potential tradeoffs coming from uncertainties in domain wall placement and diffusion will be addressed in future work. 
Here we limited our simulations to domain widths of one tenth of the coherence length.

The degree of separability of the JSA has been estimated through the Schmidt method by using the singular value decomposition \cite{law2000continuous,laudenbach2016modelling}. It's important to notice that this method strictly depends on the spectral range under examination: too small ranges lead to inaccurate results, too wide ranges are computationally intractable. It also depends on the discretisation of the JSA: too course a discretisation make the purity artificially high. An empirically-found reasonable choice for these parameters is a range of about eight times the PMF spectral width, and a JSA discretised as a $100$ by $100$ matrix.

\begin{figure*}[htb]\center
\includegraphics[width=0.85\textwidth]{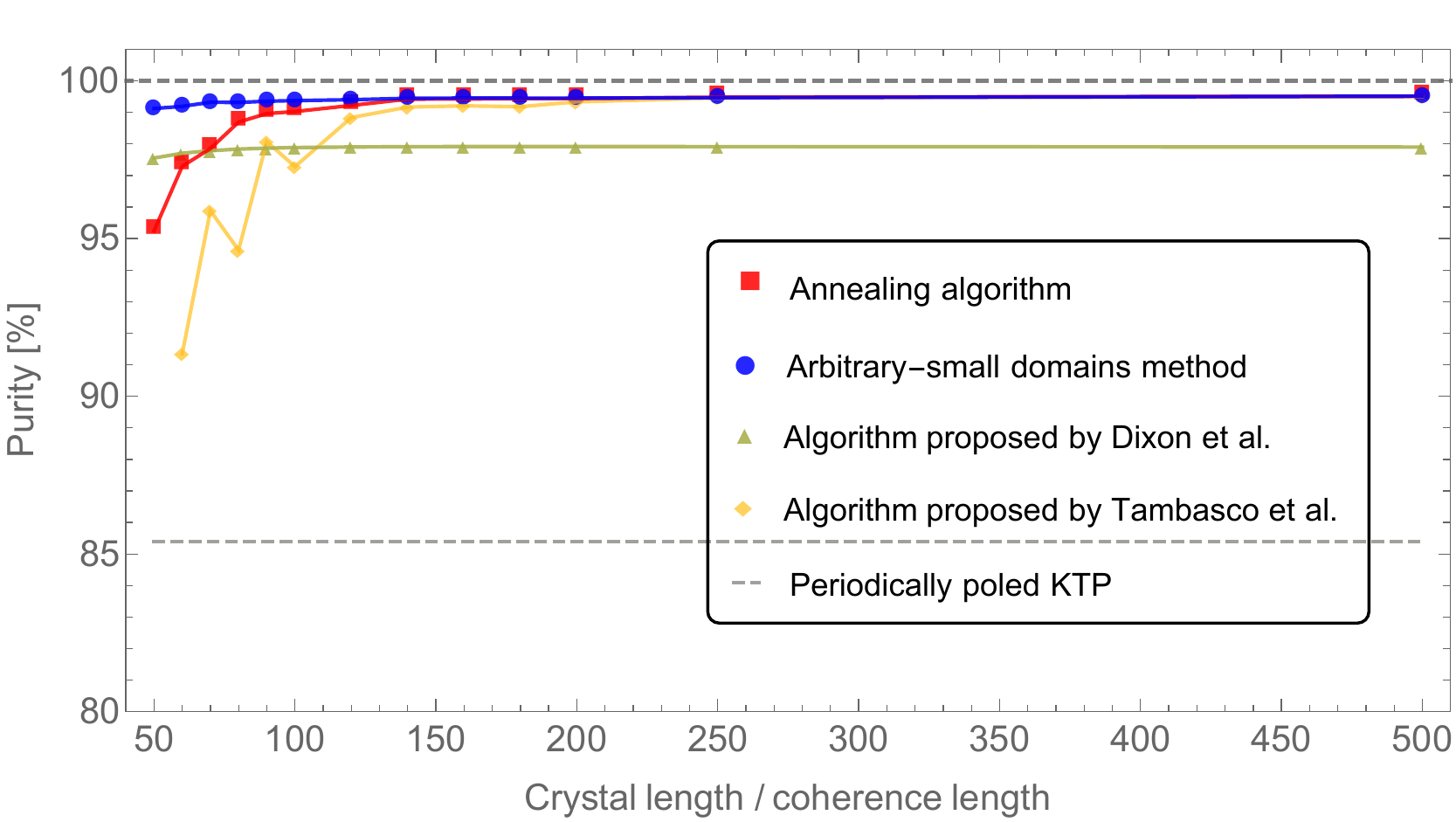}
\caption{
Purities corresponding to different lengths of KTP crystals pumped with a Gaussian pulse centred at 791 nm.
}
\label{algorithms_length}
\end{figure*}

In Fig. \ref{algorithms_length} we compare the single-photon purity for different methods---our two new methods, and those by  Dixon \emph{et al.} \cite{dixon2013spectral} and Tambasco \emph{et al.} \cite{tambasco2016domain}---as a function of crystal length (for $\sigma = L/4$). Note that we do not compare with the Dosseva \emph{et al.} method \cite{dosseva2016shaping} as this method uses simulated annealing to solve a problem to which the Tambasco \emph{et al.} method \cite{tambasco2016domain} finds the optimal solution. In other words, the purity given by \cite{dosseva2016shaping} will be bounded from above by the purity given by \cite{tambasco2016domain} for any set of parameters.

For all algorithms, the purity is highest for long crystals. The method proposed by Dixon \emph{et al.} \cite{dixon2013spectral} has an almost constant value of $97.9\%$ over the range we considered. The purity for all other methods saturates at the same value of $99.5 \%$.  Note that what limits the purity here is the choice of $\sigma$, not the algorithms themselves. This value can be increased up to $99.9 \%$ by taking $\sigma = L/5$ at the expense of a lower overall nonlinearity.

The difference between the algorithms is more drastic for short crystals. While the arbitrary-small domains method always achieves an almost optimal result, the annealing algorithm provides less separable JSAs than the algorithm proposed by Dixon \emph{et al.} \cite{dixon2013spectral} for crystals having less than $\sim 60$ domains: this might be due to a sub-optimal initial condition for the annealing process which makes the algorithm converge to a local minimum.

The unmodified Tambasco \emph{et al.} algorithm \cite{tambasco2016domain} is the only one that does not yield purities that monotonically increase with crystal length. For crystals of length $80\ell_c$ and $100\ell_c$, the purity drops when compared with slightly shorter crystals. This can be explained by considering that the two-domain-block structure of the algorithm doesn't allow enough flexibility to track the target amplitude closely due to the rough discretisation when the total number of domains is small. Numerical artefacts in the Fourier reconstruction prevented a reliable value for the purity for a $50\ell_c$ long crystal to be calculated.

It is finally  worth mentioning that the symmetrisation of our algorithms is a straightforward procedure whenever the target phase matching function is symmetric (if it is not the case, then it is not possible to have the same spectral response in both directions): either the annealing and the domain engineering with arbitrary small domains can be applied to the first half of the crystal and reproduced specularly on the second half obtaining a symmetric crystal. This could be necessary in experiments which involve a double pump configuration, where a symmetric crystal may be required for having spectrally indistinguishable photons in both directions.

\section{Conclusion}

Scalable, fault-tolerant quantum photonics will require photons of the highest purities created with near-unity heralding efficiencies. Spectral filtering severely compromises heralding efficiencies~\cite{meyer2017filtering}, increasing the number of multiplexed PDC sources required for approximating nearly-deterministic single-photon sources \cite{christ2012limits,francis2014exploring,bonneau2015effect}. At  high pump powers narrowband filtering can also  introduce undesirable mixing in other degrees of freedom \cite{branczyk2010optimized}. 

Domain engineering is therefore crucial for photonic architectures relying on parametric downconversion. The two sub-coherence length methods we introduced are fast, can readily be implemented commercially, and create higher-purity photons than previous algorithms in particular for short nonlinear crystals matched to femtosecond pump lasers~\cite{weston2016efficient}.

It will be a challenge to apply our methods to integrated photonics, where photons are typically created via four-wave mixing. Group-velocity matched four-wave mixing photon sources have been demonstrated but the observed photon purity was again limited by spurious frequency correlations caused by the sinc-shaped biphoton amplitudes~\cite{halder2009nonclassical,francis2016all}. Techniques for Gaussian apodization in four-wave-mixing have been suggested~\cite{fang2013state} but they require manipulation of the effective material dispersion which complicates simultaneous group-velocity-matching. An ongoing effort aims to introduce techniques which offer a free parameter similar to periodic poling in this platform to which we hope to map our domain-engineering algorithm.

We expect that domain engineering with arbitrarily small domains will have a range of interesting applications in 'classical' nonlinear optics for which domain engineering was originally designed~\cite{arbore1997engineerable,imeshev2000ultrashort,shur2001recent,zhang2001optimal,reid2003engineered,kornaszewski2008designer}, in particular because we can now keep track of both real and imaginary target functions which might offer advantages for phase-sensitive applications.

\vspace{0.6cm}

\begin{acknowledgments}
\noindent\textbf{Acknowledgments}

\noindent The authors acknowledge Luke G. Helt for providing valuable feedback on the manuscript. 
This work was supported by the Engineering and Physical Sciences Research Council (grant number EP/N002962/1).
F.G. acknowledges studentship funding from EPSRC under grant no. EP/L015110/1. Research at Perimeter Institute is supported by the Government of Canada through Industry Canada and by the Province of Ontario through the Ministry of Research and Innovation.
\end{acknowledgments}

%


\begin{thebibliography}{46}%
\makeatletter
\providecommand \@ifxundefined [1]{%
 \@ifx{#1\undefined}
}%
\providecommand \@ifnum [1]{%
 \ifnum #1\expandafter \@firstoftwo
 \else \expandafter \@secondoftwo
 \fi
}%
\providecommand \@ifx [1]{%
 \ifx #1\expandafter \@firstoftwo
 \else \expandafter \@secondoftwo
 \fi
}%
\providecommand \natexlab [1]{#1}%
\providecommand \enquote  [1]{``#1''}%
\providecommand \bibnamefont  [1]{#1}%
\providecommand \bibfnamefont [1]{#1}%
\providecommand \citenamefont [1]{#1}%
\providecommand \href@noop [0]{\@secondoftwo}%
\providecommand \href [0]{\begingroup \@sanitize@url \@href}%
\providecommand \@href[1]{\@@startlink{#1}\@@href}%
\providecommand \@@href[1]{\endgroup#1\@@endlink}%
\providecommand \@sanitize@url [0]{\catcode `\\12\catcode `\$12\catcode
  `\&12\catcode `\#12\catcode `\^12\catcode `\_12\catcode `\%12\relax}%
\providecommand \@@startlink[1]{}%
\providecommand \@@endlink[0]{}%
\providecommand \url  [0]{\begingroup\@sanitize@url \@url }%
\providecommand \@url [1]{\endgroup\@href {#1}{\urlprefix }}%
\providecommand \urlprefix  [0]{URL }%
\providecommand \Eprint [0]{\href }%
\providecommand \doibase [0]{http://dx.doi.org/}%
\providecommand \selectlanguage [0]{\@gobble}%
\providecommand \bibinfo  [0]{\@secondoftwo}%
\providecommand \bibfield  [0]{\@secondoftwo}%
\providecommand \translation [1]{[#1]}%
\providecommand \BibitemOpen [0]{}%
\providecommand \bibitemStop [0]{}%
\providecommand \bibitemNoStop [0]{.\EOS\space}%
\providecommand \EOS [0]{\spacefactor3000\relax}%
\providecommand \BibitemShut  [1]{\csname bibitem#1\endcsname}%
\let\auto@bib@innerbib\@empty
\bibitem [{\citenamefont {Lounis}\ and\ \citenamefont
  {Orrit}(2005)}]{lounis2005single}%
  \BibitemOpen
  \bibfield  {author} {\bibinfo {author} {\bibfnamefont {B.}~\bibnamefont
  {Lounis}}\ and\ \bibinfo {author} {\bibfnamefont {M.}~\bibnamefont {Orrit}},\
  }\href@noop {} {\bibfield  {journal} {\bibinfo  {journal} {Reports on
  Progress in Physics}\ }\textbf {\bibinfo {volume} {68}},\ \bibinfo {pages}
  {1129} (\bibinfo {year} {2005})}\BibitemShut {NoStop}%
\bibitem [{\citenamefont {Shields}(2007)}]{shields2007semiconductor}%
  \BibitemOpen
  \bibfield  {author} {\bibinfo {author} {\bibfnamefont {A.~J.}\ \bibnamefont
  {Shields}},\ }\href@noop {} {\bibfield  {journal} {\bibinfo  {journal}
  {Nature Photonics}\ }\textbf {\bibinfo {volume} {1}},\ \bibinfo {pages} {215}
  (\bibinfo {year} {2007})}\BibitemShut {NoStop}%
\bibitem [{\citenamefont {Michler}(2009)}]{Michler2009}%
  \BibitemOpen
  \bibfield  {author} {\bibinfo {author} {\bibfnamefont {P.}~\bibnamefont
  {Michler}},\ }\enquote {\bibinfo {title} {Quantum dot single-photon
  sources},}\ in\ \href {\doibase 10.1007/978-3-540-87446-1_6} {\emph {\bibinfo
  {booktitle} {Single Semiconductor Quantum Dots}}},\ \bibinfo {editor} {edited
  by\ \bibinfo {editor} {\bibfnamefont {P.}~\bibnamefont {Michler}}}\ (\bibinfo
   {publisher} {Springer Berlin Heidelberg},\ \bibinfo {address} {Berlin,
  Heidelberg},\ \bibinfo {year} {2009})\ pp.\ \bibinfo {pages}
  {185--225}\BibitemShut {NoStop}%
\bibitem [{\citenamefont {Burnham}\ and\ \citenamefont
  {Weinberg}(1970)}]{burnham1970observation}%
  \BibitemOpen
  \bibfield  {author} {\bibinfo {author} {\bibfnamefont {D.~C.}\ \bibnamefont
  {Burnham}}\ and\ \bibinfo {author} {\bibfnamefont {D.~L.}\ \bibnamefont
  {Weinberg}},\ }\href@noop {} {\bibfield  {journal} {\bibinfo  {journal}
  {Physical Review Letters}\ }\textbf {\bibinfo {volume} {25}},\ \bibinfo
  {pages} {84} (\bibinfo {year} {1970})}\BibitemShut {NoStop}%
\bibitem [{\citenamefont {Kwiat}\ \emph {et~al.}(1995)\citenamefont {Kwiat},
  \citenamefont {Mattle}, \citenamefont {Weinfurter}, \citenamefont
  {Zeilinger}, \citenamefont {Sergienko},\ and\ \citenamefont
  {Shih}}]{kwiat1995new}%
  \BibitemOpen
  \bibfield  {author} {\bibinfo {author} {\bibfnamefont {P.~G.}\ \bibnamefont
  {Kwiat}}, \bibinfo {author} {\bibfnamefont {K.}~\bibnamefont {Mattle}},
  \bibinfo {author} {\bibfnamefont {H.}~\bibnamefont {Weinfurter}}, \bibinfo
  {author} {\bibfnamefont {A.}~\bibnamefont {Zeilinger}}, \bibinfo {author}
  {\bibfnamefont {A.~V.}\ \bibnamefont {Sergienko}}, \ and\ \bibinfo {author}
  {\bibfnamefont {Y.}~\bibnamefont {Shih}},\ }\href@noop {} {\bibfield
  {journal} {\bibinfo  {journal} {Physical Review Letters}\ }\textbf {\bibinfo
  {volume} {75}},\ \bibinfo {pages} {4337} (\bibinfo {year}
  {1995})}\BibitemShut {NoStop}%
\bibitem [{\citenamefont {Tanzilli}\ \emph {et~al.}(2001)\citenamefont
  {Tanzilli}, \citenamefont {De~Riedmatten}, \citenamefont {Tittel},
  \citenamefont {Zbinden}, \citenamefont {Baldi}, \citenamefont {De~Micheli},
  \citenamefont {Ostrowsky},\ and\ \citenamefont {Gisin}}]{tanzilli2001highly}%
  \BibitemOpen
  \bibfield  {author} {\bibinfo {author} {\bibfnamefont {S.}~\bibnamefont
  {Tanzilli}}, \bibinfo {author} {\bibfnamefont {H.}~\bibnamefont
  {De~Riedmatten}}, \bibinfo {author} {\bibfnamefont {H.}~\bibnamefont
  {Tittel}}, \bibinfo {author} {\bibfnamefont {H.}~\bibnamefont {Zbinden}},
  \bibinfo {author} {\bibfnamefont {P.}~\bibnamefont {Baldi}}, \bibinfo
  {author} {\bibfnamefont {M.}~\bibnamefont {De~Micheli}}, \bibinfo {author}
  {\bibfnamefont {D.~B.}\ \bibnamefont {Ostrowsky}}, \ and\ \bibinfo {author}
  {\bibfnamefont {N.}~\bibnamefont {Gisin}},\ }\href@noop {} {\bibfield
  {journal} {\bibinfo  {journal} {Electronics Letters}\ }\textbf {\bibinfo
  {volume} {37}},\ \bibinfo {pages} {26} (\bibinfo {year} {2001})}\BibitemShut
  {NoStop}%
\bibitem [{\citenamefont {Mosley}\ \emph {et~al.}(2008)\citenamefont {Mosley},
  \citenamefont {Lundeen}, \citenamefont {Smith}, \citenamefont {Wasylczyk},
  \citenamefont {U'Ren}, \citenamefont {Silberhorn},\ and\ \citenamefont
  {Walmsley}}]{mosley2008heralded}%
  \BibitemOpen
  \bibfield  {author} {\bibinfo {author} {\bibfnamefont {P.~J.}\ \bibnamefont
  {Mosley}}, \bibinfo {author} {\bibfnamefont {J.~S.}\ \bibnamefont {Lundeen}},
  \bibinfo {author} {\bibfnamefont {B.~J.}\ \bibnamefont {Smith}}, \bibinfo
  {author} {\bibfnamefont {P.}~\bibnamefont {Wasylczyk}}, \bibinfo {author}
  {\bibfnamefont {A.~B.}\ \bibnamefont {U'Ren}}, \bibinfo {author}
  {\bibfnamefont {C.}~\bibnamefont {Silberhorn}}, \ and\ \bibinfo {author}
  {\bibfnamefont {I.~A.}\ \bibnamefont {Walmsley}},\ }\href@noop {} {\bibfield
  {journal} {\bibinfo  {journal} {Physical Review Letters}\ }\textbf {\bibinfo
  {volume} {100}},\ \bibinfo {pages} {133601} (\bibinfo {year}
  {2008})}\BibitemShut {NoStop}%
\bibitem [{\citenamefont {Jennewein}\ \emph {et~al.}(2011)\citenamefont
  {Jennewein}, \citenamefont {Barbieri},\ and\ \citenamefont
  {White}}]{jennewein2011single}%
  \BibitemOpen
  \bibfield  {author} {\bibinfo {author} {\bibfnamefont {T.}~\bibnamefont
  {Jennewein}}, \bibinfo {author} {\bibfnamefont {M.}~\bibnamefont {Barbieri}},
  \ and\ \bibinfo {author} {\bibfnamefont {A.~G.}\ \bibnamefont {White}},\
  }\href@noop {} {\bibfield  {journal} {\bibinfo  {journal} {Journal of Modern
  Optics}\ }\textbf {\bibinfo {volume} {58}},\ \bibinfo {pages} {276} (\bibinfo
  {year} {2011})}\BibitemShut {NoStop}%
\bibitem [{\citenamefont {Christ}\ and\ \citenamefont
  {Silberhorn}(2012)}]{christ2012limits}%
  \BibitemOpen
  \bibfield  {author} {\bibinfo {author} {\bibfnamefont {A.}~\bibnamefont
  {Christ}}\ and\ \bibinfo {author} {\bibfnamefont {C.}~\bibnamefont
  {Silberhorn}},\ }\href@noop {} {\bibfield  {journal} {\bibinfo  {journal}
  {Physical Review A}\ }\textbf {\bibinfo {volume} {85}},\ \bibinfo {pages}
  {023829} (\bibinfo {year} {2012})}\BibitemShut {NoStop}%
\bibitem [{\citenamefont {Meyer-Scott}\ \emph {et~al.}(2017)\citenamefont
  {Meyer-Scott}, \citenamefont {Montaut}, \citenamefont {Tiedau}, \citenamefont
  {Sansoni}, \citenamefont {Herrmann}, \citenamefont {Bartley},\ and\
  \citenamefont {Silberhorn}}]{meyer2017filtering}%
  \BibitemOpen
  \bibfield  {author} {\bibinfo {author} {\bibfnamefont {E.}~\bibnamefont
  {Meyer-Scott}}, \bibinfo {author} {\bibfnamefont {N.}~\bibnamefont
  {Montaut}}, \bibinfo {author} {\bibfnamefont {J.}~\bibnamefont {Tiedau}},
  \bibinfo {author} {\bibfnamefont {L.}~\bibnamefont {Sansoni}}, \bibinfo
  {author} {\bibfnamefont {H.}~\bibnamefont {Herrmann}}, \bibinfo {author}
  {\bibfnamefont {T.~J.}\ \bibnamefont {Bartley}}, \ and\ \bibinfo {author}
  {\bibfnamefont {C.}~\bibnamefont {Silberhorn}},\ }\href@noop {} {\bibfield
  {journal} {\bibinfo  {journal} {arXiv preprint arXiv:1702.05501}\ } (\bibinfo
  {year} {2017})}\BibitemShut {NoStop}%
\bibitem [{\citenamefont {Bra{\'n}czyk}\ \emph {et~al.}(2010)\citenamefont
  {Bra{\'n}czyk}, \citenamefont {Ralph}, \citenamefont {Helwig},\ and\
  \citenamefont {Silberhorn}}]{branczyk2010optimized}%
  \BibitemOpen
  \bibfield  {author} {\bibinfo {author} {\bibfnamefont {A.~M.}\ \bibnamefont
  {Bra{\'n}czyk}}, \bibinfo {author} {\bibfnamefont {T.}~\bibnamefont {Ralph}},
  \bibinfo {author} {\bibfnamefont {W.}~\bibnamefont {Helwig}}, \ and\ \bibinfo
  {author} {\bibfnamefont {C.}~\bibnamefont {Silberhorn}},\ }\href@noop {}
  {\bibfield  {journal} {\bibinfo  {journal} {New Journal of Physics}\ }\textbf
  {\bibinfo {volume} {12}},\ \bibinfo {pages} {063001} (\bibinfo {year}
  {2010})}\BibitemShut {NoStop}%
\bibitem [{\citenamefont {Keller}\ and\ \citenamefont
  {Rubin}(1997)}]{keller1997theory}%
  \BibitemOpen
  \bibfield  {author} {\bibinfo {author} {\bibfnamefont {T.~E.}\ \bibnamefont
  {Keller}}\ and\ \bibinfo {author} {\bibfnamefont {M.~H.}\ \bibnamefont
  {Rubin}},\ }\href@noop {} {\bibfield  {journal} {\bibinfo  {journal}
  {Physical Review A}\ }\textbf {\bibinfo {volume} {56}},\ \bibinfo {pages}
  {1534} (\bibinfo {year} {1997})}\BibitemShut {NoStop}%
\bibitem [{\citenamefont {Grice}\ \emph {et~al.}(2001)\citenamefont {Grice},
  \citenamefont {U'Ren},\ and\ \citenamefont
  {Walmsley}}]{grice2001eliminating}%
  \BibitemOpen
  \bibfield  {author} {\bibinfo {author} {\bibfnamefont {W.}~\bibnamefont
  {Grice}}, \bibinfo {author} {\bibfnamefont {A.}~\bibnamefont {U'Ren}}, \ and\
  \bibinfo {author} {\bibfnamefont {I.}~\bibnamefont {Walmsley}},\ }\href@noop
  {} {\bibfield  {journal} {\bibinfo  {journal} {Physical Review A}\ }\textbf
  {\bibinfo {volume} {64}},\ \bibinfo {pages} {063815} (\bibinfo {year}
  {2001})}\BibitemShut {NoStop}%
\bibitem [{\citenamefont {U'Ren}\ \emph {et~al.}(2006)\citenamefont {U'Ren},
  \citenamefont {Silberhorn}, \citenamefont {Erdmann}, \citenamefont
  {Banaszek}, \citenamefont {Grice}, \citenamefont {Walmsley},\ and\
  \citenamefont {Raymer}}]{u2006generation}%
  \BibitemOpen
  \bibfield  {author} {\bibinfo {author} {\bibfnamefont {A.~B.}\ \bibnamefont
  {U'Ren}}, \bibinfo {author} {\bibfnamefont {C.}~\bibnamefont {Silberhorn}},
  \bibinfo {author} {\bibfnamefont {R.}~\bibnamefont {Erdmann}}, \bibinfo
  {author} {\bibfnamefont {K.}~\bibnamefont {Banaszek}}, \bibinfo {author}
  {\bibfnamefont {W.~P.}\ \bibnamefont {Grice}}, \bibinfo {author}
  {\bibfnamefont {I.~A.}\ \bibnamefont {Walmsley}}, \ and\ \bibinfo {author}
  {\bibfnamefont {M.~G.}\ \bibnamefont {Raymer}},\ }\href@noop {} {\bibfield
  {journal} {\bibinfo  {journal} {arXiv preprint quant-ph/0611019}\ } (\bibinfo
  {year} {2006})}\BibitemShut {NoStop}%
\bibitem [{\citenamefont {Fejer}\ \emph {et~al.}(1992)\citenamefont {Fejer},
  \citenamefont {Magel}, \citenamefont {Jundt},\ and\ \citenamefont
  {Byer}}]{fejer1992quasi}%
  \BibitemOpen
  \bibfield  {author} {\bibinfo {author} {\bibfnamefont {M.~M.}\ \bibnamefont
  {Fejer}}, \bibinfo {author} {\bibfnamefont {G.}~\bibnamefont {Magel}},
  \bibinfo {author} {\bibfnamefont {D.~H.}\ \bibnamefont {Jundt}}, \ and\
  \bibinfo {author} {\bibfnamefont {R.~L.}\ \bibnamefont {Byer}},\ }\href@noop
  {} {\bibfield  {journal} {\bibinfo  {journal} {IEEE Journal of Quantum
  Electronics}\ }\textbf {\bibinfo {volume} {28}},\ \bibinfo {pages} {2631}
  (\bibinfo {year} {1992})}\BibitemShut {NoStop}%
\bibitem [{\citenamefont {Bra{\'n}czyk}\ \emph
  {et~al.}(2011{\natexlab{a}})\citenamefont {Bra{\'n}czyk}, \citenamefont
  {Fedrizzi}, \citenamefont {Stace}, \citenamefont {Ralph},\ and\ \citenamefont
  {White}}]{branczyk2011engineered}%
  \BibitemOpen
  \bibfield  {author} {\bibinfo {author} {\bibfnamefont {A.~M.}\ \bibnamefont
  {Bra{\'n}czyk}}, \bibinfo {author} {\bibfnamefont {A.}~\bibnamefont
  {Fedrizzi}}, \bibinfo {author} {\bibfnamefont {T.~M.}\ \bibnamefont {Stace}},
  \bibinfo {author} {\bibfnamefont {T.~C.}\ \bibnamefont {Ralph}}, \ and\
  \bibinfo {author} {\bibfnamefont {A.~G.}\ \bibnamefont {White}},\ }\href@noop
  {} {\bibfield  {journal} {\bibinfo  {journal} {Optics Express}\ }\textbf
  {\bibinfo {volume} {19}},\ \bibinfo {pages} {55} (\bibinfo {year}
  {2011}{\natexlab{a}})}\BibitemShut {NoStop}%
\bibitem [{\citenamefont {Dixon}\ \emph {et~al.}(2013)\citenamefont {Dixon},
  \citenamefont {Shapiro},\ and\ \citenamefont {Wong}}]{dixon2013spectral}%
  \BibitemOpen
  \bibfield  {author} {\bibinfo {author} {\bibfnamefont {P.~B.}\ \bibnamefont
  {Dixon}}, \bibinfo {author} {\bibfnamefont {J.~H.}\ \bibnamefont {Shapiro}},
  \ and\ \bibinfo {author} {\bibfnamefont {F.~N.}\ \bibnamefont {Wong}},\
  }\href@noop {} {\bibfield  {journal} {\bibinfo  {journal} {Optics Express}\
  }\textbf {\bibinfo {volume} {21}},\ \bibinfo {pages} {5879} (\bibinfo {year}
  {2013})}\BibitemShut {NoStop}%
\bibitem [{\citenamefont {Dosseva}\ \emph {et~al.}(2016)\citenamefont
  {Dosseva}, \citenamefont {Cincio},\ and\ \citenamefont
  {Bra{\'n}czyk}}]{dosseva2016shaping}%
  \BibitemOpen
  \bibfield  {author} {\bibinfo {author} {\bibfnamefont {A.}~\bibnamefont
  {Dosseva}}, \bibinfo {author} {\bibfnamefont {{\L}.}~\bibnamefont {Cincio}},
  \ and\ \bibinfo {author} {\bibfnamefont {A.~M.}\ \bibnamefont
  {Bra{\'n}czyk}},\ }\href@noop {} {\bibfield  {journal} {\bibinfo  {journal}
  {Physical Review A}\ }\textbf {\bibinfo {volume} {93}},\ \bibinfo {pages}
  {013801} (\bibinfo {year} {2016})}\BibitemShut {NoStop}%
\bibitem [{\citenamefont {Tambasco}\ \emph {et~al.}(2016)\citenamefont
  {Tambasco}, \citenamefont {Boes}, \citenamefont {Helt}, \citenamefont
  {Steel},\ and\ \citenamefont {Mitchell}}]{tambasco2016domain}%
  \BibitemOpen
  \bibfield  {author} {\bibinfo {author} {\bibfnamefont {J.}~\bibnamefont
  {Tambasco}}, \bibinfo {author} {\bibfnamefont {A.}~\bibnamefont {Boes}},
  \bibinfo {author} {\bibfnamefont {L.}~\bibnamefont {Helt}}, \bibinfo {author}
  {\bibfnamefont {M.}~\bibnamefont {Steel}}, \ and\ \bibinfo {author}
  {\bibfnamefont {A.}~\bibnamefont {Mitchell}},\ }\href@noop {} {\bibfield
  {journal} {\bibinfo  {journal} {Optics Express}\ }\textbf {\bibinfo {volume}
  {24}},\ \bibinfo {pages} {19616} (\bibinfo {year} {2016})}\BibitemShut
  {NoStop}%
\bibitem [{\citenamefont {Chen}\ \emph {et~al.}(2017)\citenamefont {Chen},
  \citenamefont {Bo}, \citenamefont {Niu}, \citenamefont {Xu}, \citenamefont
  {Zhang}, \citenamefont {Shapiro},\ and\ \citenamefont
  {Wong}}]{chen2017efficient}%
  \BibitemOpen
  \bibfield  {author} {\bibinfo {author} {\bibfnamefont {C.}~\bibnamefont
  {Chen}}, \bibinfo {author} {\bibfnamefont {C.}~\bibnamefont {Bo}}, \bibinfo
  {author} {\bibfnamefont {M.~Y.}\ \bibnamefont {Niu}}, \bibinfo {author}
  {\bibfnamefont {F.}~\bibnamefont {Xu}}, \bibinfo {author} {\bibfnamefont
  {Z.}~\bibnamefont {Zhang}}, \bibinfo {author} {\bibfnamefont {J.~H.}\
  \bibnamefont {Shapiro}}, \ and\ \bibinfo {author} {\bibfnamefont {F.~N.}\
  \bibnamefont {Wong}},\ }\href@noop {} {\bibfield  {journal} {\bibinfo
  {journal} {Optics Express}\ }\textbf {\bibinfo {volume} {25}},\ \bibinfo
  {pages} {7300} (\bibinfo {year} {2017})}\BibitemShut {NoStop}%
\bibitem [{\citenamefont {Hong}\ \emph {et~al.}(1987)\citenamefont {Hong},
  \citenamefont {Ou},\ and\ \citenamefont {Mandel}}]{hong1987measurement}%
  \BibitemOpen
  \bibfield  {author} {\bibinfo {author} {\bibfnamefont {C.}~\bibnamefont
  {Hong}}, \bibinfo {author} {\bibfnamefont {Z.}~\bibnamefont {Ou}}, \ and\
  \bibinfo {author} {\bibfnamefont {L.}~\bibnamefont {Mandel}},\ }\href@noop {}
  {\bibfield  {journal} {\bibinfo  {journal} {Physical Review Letters}\
  }\textbf {\bibinfo {volume} {59}},\ \bibinfo {pages} {2044} (\bibinfo {year}
  {1987})}\BibitemShut {NoStop}%
\bibitem [{\citenamefont {Reid}(2003)}]{reid2003engineered}%
  \BibitemOpen
  \bibfield  {author} {\bibinfo {author} {\bibfnamefont {D.~T.}\ \bibnamefont
  {Reid}},\ }\href@noop {} {\bibfield  {journal} {\bibinfo  {journal} {Journal
  of Optics A: Pure and Applied Optics}\ }\textbf {\bibinfo {volume} {5}},\
  \bibinfo {pages} {S97} (\bibinfo {year} {2003})}\BibitemShut {NoStop}%
\bibitem [{\citenamefont {Kornaszewski}\ \emph {et~al.}(2008)\citenamefont
  {Kornaszewski}, \citenamefont {Kohler}, \citenamefont {Sapaev},\ and\
  \citenamefont {Reid}}]{kornaszewski2008designer}%
  \BibitemOpen
  \bibfield  {author} {\bibinfo {author} {\bibfnamefont {{\L}.}~\bibnamefont
  {Kornaszewski}}, \bibinfo {author} {\bibfnamefont {M.}~\bibnamefont
  {Kohler}}, \bibinfo {author} {\bibfnamefont {U.~K.}\ \bibnamefont {Sapaev}},
  \ and\ \bibinfo {author} {\bibfnamefont {D.~T.}\ \bibnamefont {Reid}},\
  }\href@noop {} {\bibfield  {journal} {\bibinfo  {journal} {Optics Letters}\
  }\textbf {\bibinfo {volume} {33}},\ \bibinfo {pages} {378} (\bibinfo {year}
  {2008})}\BibitemShut {NoStop}%
\bibitem [{\citenamefont {Harder}(2016)}]{harder2016optimized}%
  \BibitemOpen
  \bibfield  {author} {\bibinfo {author} {\bibfnamefont {G.}~\bibnamefont
  {Harder}},\ }\emph {\bibinfo {title} {Optimized down-conversion source and
  state-characterization tools for quantum optics}},\ \href@noop {} {Ph.D.
  thesis},\ \bibinfo  {school} {Dissertation, Paderborn, Universit{\"a}t
  Paderborn, 2016} (\bibinfo {year} {2016})\BibitemShut {NoStop}%
\bibitem [{\citenamefont {Bra{\'n}czyk}\ \emph
  {et~al.}(2011{\natexlab{b}})\citenamefont {Bra{\'n}czyk}, \citenamefont
  {Stace},\ and\ \citenamefont {Ralph}}]{branczyk2011time}%
  \BibitemOpen
  \bibfield  {author} {\bibinfo {author} {\bibfnamefont {A.~M.}\ \bibnamefont
  {Bra{\'n}czyk}}, \bibinfo {author} {\bibfnamefont {T.~M.}\ \bibnamefont
  {Stace}}, \ and\ \bibinfo {author} {\bibfnamefont {T.~C.}\ \bibnamefont
  {Ralph}},\ }in\ \href@noop {} {\emph {\bibinfo {booktitle} {AIP Conference
  Proceedings}}},\ Vol.\ \bibinfo {volume} {1363}\ (\bibinfo {organization}
  {AIP},\ \bibinfo {year} {2011})\ pp.\ \bibinfo {pages} {335--338}\BibitemShut
  {NoStop}%
\bibitem [{\citenamefont {Christ}\ \emph {et~al.}(2013)\citenamefont {Christ},
  \citenamefont {Brecht}, \citenamefont {Mauerer},\ and\ \citenamefont
  {Silberhorn}}]{christ2013theory}%
  \BibitemOpen
  \bibfield  {author} {\bibinfo {author} {\bibfnamefont {A.}~\bibnamefont
  {Christ}}, \bibinfo {author} {\bibfnamefont {B.}~\bibnamefont {Brecht}},
  \bibinfo {author} {\bibfnamefont {W.}~\bibnamefont {Mauerer}}, \ and\
  \bibinfo {author} {\bibfnamefont {C.}~\bibnamefont {Silberhorn}},\
  }\href@noop {} {\bibfield  {journal} {\bibinfo  {journal} {New Journal of
  Physics}\ }\textbf {\bibinfo {volume} {15}},\ \bibinfo {pages} {053038}
  (\bibinfo {year} {2013})}\BibitemShut {NoStop}%
\bibitem [{\citenamefont {Quesada}\ and\ \citenamefont
  {Sipe}(2014)}]{quesada2014effects}%
  \BibitemOpen
  \bibfield  {author} {\bibinfo {author} {\bibfnamefont {N.}~\bibnamefont
  {Quesada}}\ and\ \bibinfo {author} {\bibfnamefont {J.}~\bibnamefont {Sipe}},\
  }\href@noop {} {\bibfield  {journal} {\bibinfo  {journal} {Physical Review
  A}\ }\textbf {\bibinfo {volume} {90}},\ \bibinfo {pages} {063840} (\bibinfo
  {year} {2014})}\BibitemShut {NoStop}%
\bibitem [{\citenamefont {Quesada}\ and\ \citenamefont
  {Sipe}(2015)}]{quesada2015time}%
  \BibitemOpen
  \bibfield  {author} {\bibinfo {author} {\bibfnamefont {N.}~\bibnamefont
  {Quesada}}\ and\ \bibinfo {author} {\bibfnamefont {J.}~\bibnamefont {Sipe}},\
  }\href@noop {} {\bibfield  {journal} {\bibinfo  {journal} {Physical Review
  Letters}\ }\textbf {\bibinfo {volume} {114}},\ \bibinfo {pages} {093903}
  (\bibinfo {year} {2015})}\BibitemShut {NoStop}%
\bibitem [{\citenamefont {Laudenbach}\ \emph {et~al.}(2016)\citenamefont
  {Laudenbach}, \citenamefont {H{\"u}bel}, \citenamefont {Hentschel},
  \citenamefont {Walther},\ and\ \citenamefont
  {Poppe}}]{laudenbach2016modelling}%
  \BibitemOpen
  \bibfield  {author} {\bibinfo {author} {\bibfnamefont {F.}~\bibnamefont
  {Laudenbach}}, \bibinfo {author} {\bibfnamefont {H.}~\bibnamefont
  {H{\"u}bel}}, \bibinfo {author} {\bibfnamefont {M.}~\bibnamefont
  {Hentschel}}, \bibinfo {author} {\bibfnamefont {P.}~\bibnamefont {Walther}},
  \ and\ \bibinfo {author} {\bibfnamefont {A.}~\bibnamefont {Poppe}},\
  }\href@noop {} {\bibfield  {journal} {\bibinfo  {journal} {Optics Express}\
  }\textbf {\bibinfo {volume} {24}},\ \bibinfo {pages} {2712} (\bibinfo {year}
  {2016})}\BibitemShut {NoStop}%
\bibitem [{\citenamefont {Arbore}\ \emph {et~al.}(1997)\citenamefont {Arbore},
  \citenamefont {Galvanauskas}, \citenamefont {Harter}, \citenamefont {Chou},\
  and\ \citenamefont {Fejer}}]{arbore1997engineerable}%
  \BibitemOpen
  \bibfield  {author} {\bibinfo {author} {\bibfnamefont {M.}~\bibnamefont
  {Arbore}}, \bibinfo {author} {\bibfnamefont {A.}~\bibnamefont
  {Galvanauskas}}, \bibinfo {author} {\bibfnamefont {D.}~\bibnamefont
  {Harter}}, \bibinfo {author} {\bibfnamefont {M.}~\bibnamefont {Chou}}, \ and\
  \bibinfo {author} {\bibfnamefont {M.}~\bibnamefont {Fejer}},\ }\href@noop {}
  {\bibfield  {journal} {\bibinfo  {journal} {Optics Letters}\ }\textbf
  {\bibinfo {volume} {22}},\ \bibinfo {pages} {1341} (\bibinfo {year}
  {1997})}\BibitemShut {NoStop}%
\bibitem [{\citenamefont {Imeshev}\ \emph {et~al.}(2000)\citenamefont
  {Imeshev}, \citenamefont {Arbore}, \citenamefont {Fejer}, \citenamefont
  {Galvanauskas}, \citenamefont {Fermann},\ and\ \citenamefont
  {Harter}}]{imeshev2000ultrashort}%
  \BibitemOpen
  \bibfield  {author} {\bibinfo {author} {\bibfnamefont {G.}~\bibnamefont
  {Imeshev}}, \bibinfo {author} {\bibfnamefont {M.}~\bibnamefont {Arbore}},
  \bibinfo {author} {\bibfnamefont {M.}~\bibnamefont {Fejer}}, \bibinfo
  {author} {\bibfnamefont {A.}~\bibnamefont {Galvanauskas}}, \bibinfo {author}
  {\bibfnamefont {M.}~\bibnamefont {Fermann}}, \ and\ \bibinfo {author}
  {\bibfnamefont {D.}~\bibnamefont {Harter}},\ }\href@noop {} {\bibfield
  {journal} {\bibinfo  {journal} {JOSA B}\ }\textbf {\bibinfo {volume} {17}},\
  \bibinfo {pages} {304} (\bibinfo {year} {2000})}\BibitemShut {NoStop}%
\bibitem [{\citenamefont {Shur}\ \emph {et~al.}(2001)\citenamefont {Shur},
  \citenamefont {Rumyantsev}, \citenamefont {Ndcolaeva}, \citenamefont
  {Shishkin}, \citenamefont {Batchko}, \citenamefont {Fejer},\ and\
  \citenamefont {Byer}}]{shur2001recent}%
  \BibitemOpen
  \bibfield  {author} {\bibinfo {author} {\bibfnamefont {V.~Y.}\ \bibnamefont
  {Shur}}, \bibinfo {author} {\bibfnamefont {E.}~\bibnamefont {Rumyantsev}},
  \bibinfo {author} {\bibfnamefont {E.}~\bibnamefont {Ndcolaeva}}, \bibinfo
  {author} {\bibfnamefont {E.}~\bibnamefont {Shishkin}}, \bibinfo {author}
  {\bibfnamefont {R.}~\bibnamefont {Batchko}}, \bibinfo {author} {\bibfnamefont
  {M.}~\bibnamefont {Fejer}}, \ and\ \bibinfo {author} {\bibfnamefont
  {R.}~\bibnamefont {Byer}},\ }\href@noop {} {\bibfield  {journal} {\bibinfo
  {journal} {Ferroelectrics}\ }\textbf {\bibinfo {volume} {257}},\ \bibinfo
  {pages} {191} (\bibinfo {year} {2001})}\BibitemShut {NoStop}%
\bibitem [{\citenamefont {Zhang}\ and\ \citenamefont
  {Gu}(2001)}]{zhang2001optimal}%
  \BibitemOpen
  \bibfield  {author} {\bibinfo {author} {\bibfnamefont {Y.}~\bibnamefont
  {Zhang}}\ and\ \bibinfo {author} {\bibfnamefont {B.-Y.}\ \bibnamefont {Gu}},\
  }\href@noop {} {\bibfield  {journal} {\bibinfo  {journal} {Optics
  communications}\ }\textbf {\bibinfo {volume} {192}},\ \bibinfo {pages} {417}
  (\bibinfo {year} {2001})}\BibitemShut {NoStop}%
\bibitem [{\citenamefont {Boyd}(2003)}]{boyd2003nonlinear}%
  \BibitemOpen
  \bibfield  {author} {\bibinfo {author} {\bibfnamefont {R.~W.}\ \bibnamefont
  {Boyd}},\ }in\ \href@noop {} {\emph {\bibinfo {booktitle} {Handbook of Laser
  Technology and Applications (Three-Volume Set)}}}\ (\bibinfo  {publisher}
  {Taylor \& Francis},\ \bibinfo {year} {2003})\ pp.\ \bibinfo {pages}
  {161--183}\BibitemShut {NoStop}%
\bibitem [{\citenamefont {Kirkpatrick}\ \emph {et~al.}(1983)\citenamefont
  {Kirkpatrick}, \citenamefont {Gelatt}, \citenamefont {Vecchi} \emph
  {et~al.}}]{annealing1983gelatt}%
  \BibitemOpen
  \bibfield  {author} {\bibinfo {author} {\bibfnamefont {S.}~\bibnamefont
  {Kirkpatrick}}, \bibinfo {author} {\bibfnamefont {C.~D.}\ \bibnamefont
  {Gelatt}}, \bibinfo {author} {\bibfnamefont {M.~P.}\ \bibnamefont {Vecchi}},
  \emph {et~al.},\ }\href@noop {} {\bibfield  {journal} {\bibinfo  {journal}
  {Science}\ }\textbf {\bibinfo {volume} {220}},\ \bibinfo {pages} {671}
  (\bibinfo {year} {1983})}\BibitemShut {NoStop}%
\bibitem [{\citenamefont {Kirkpatrick}(1984)}]{kirkpatrick1984optimization}%
  \BibitemOpen
  \bibfield  {author} {\bibinfo {author} {\bibfnamefont {S.}~\bibnamefont
  {Kirkpatrick}},\ }\href@noop {} {\bibfield  {journal} {\bibinfo  {journal}
  {Journal of statistical physics}\ }\textbf {\bibinfo {volume} {34}},\
  \bibinfo {pages} {975} (\bibinfo {year} {1984})}\BibitemShut {NoStop}%
\bibitem [{\citenamefont {K{\"o}nig}\ and\ \citenamefont
  {Wong}(2004)}]{konig2004extended}%
  \BibitemOpen
  \bibfield  {author} {\bibinfo {author} {\bibfnamefont {F.}~\bibnamefont
  {K{\"o}nig}}\ and\ \bibinfo {author} {\bibfnamefont {F.~N.}\ \bibnamefont
  {Wong}},\ }\href@noop {} {\bibfield  {journal} {\bibinfo  {journal} {Applied
  Physics Letters}\ }\textbf {\bibinfo {volume} {84}},\ \bibinfo {pages} {1644}
  (\bibinfo {year} {2004})}\BibitemShut {NoStop}%
\bibitem [{\citenamefont {Fradkin}\ \emph {et~al.}(1999)\citenamefont
  {Fradkin}, \citenamefont {Arie}, \citenamefont {Skliar},\ and\ \citenamefont
  {Rosenman}}]{fradkin1999tunable}%
  \BibitemOpen
  \bibfield  {author} {\bibinfo {author} {\bibfnamefont {K.}~\bibnamefont
  {Fradkin}}, \bibinfo {author} {\bibfnamefont {A.}~\bibnamefont {Arie}},
  \bibinfo {author} {\bibfnamefont {A.}~\bibnamefont {Skliar}}, \ and\ \bibinfo
  {author} {\bibfnamefont {G.}~\bibnamefont {Rosenman}},\ }\href@noop {}
  {\bibfield  {journal} {\bibinfo  {journal} {Applied Physics Letters}\
  }\textbf {\bibinfo {volume} {74}},\ \bibinfo {pages} {914} (\bibinfo {year}
  {1999})}\BibitemShut {NoStop}%
\bibitem [{\citenamefont {Emanueli}\ and\ \citenamefont
  {Arie}(2003)}]{emanueli2003temperature}%
  \BibitemOpen
  \bibfield  {author} {\bibinfo {author} {\bibfnamefont {S.}~\bibnamefont
  {Emanueli}}\ and\ \bibinfo {author} {\bibfnamefont {A.}~\bibnamefont
  {Arie}},\ }\href@noop {} {\bibfield  {journal} {\bibinfo  {journal} {Applied
  Optics}\ }\textbf {\bibinfo {volume} {42}},\ \bibinfo {pages} {6661}
  (\bibinfo {year} {2003})}\BibitemShut {NoStop}%
\bibitem [{\citenamefont {Law}\ \emph {et~al.}(2000)\citenamefont {Law},
  \citenamefont {Walmsley},\ and\ \citenamefont {Eberly}}]{law2000continuous}%
  \BibitemOpen
  \bibfield  {author} {\bibinfo {author} {\bibfnamefont {C.}~\bibnamefont
  {Law}}, \bibinfo {author} {\bibfnamefont {I.}~\bibnamefont {Walmsley}}, \
  and\ \bibinfo {author} {\bibfnamefont {J.}~\bibnamefont {Eberly}},\
  }\href@noop {} {\bibfield  {journal} {\bibinfo  {journal} {Physical Review
  Letters}\ }\textbf {\bibinfo {volume} {84}},\ \bibinfo {pages} {5304}
  (\bibinfo {year} {2000})}\BibitemShut {NoStop}%
\bibitem [{\citenamefont {Francis-Jones}\ and\ \citenamefont
  {Mosley}(2014)}]{francis2014exploring}%
  \BibitemOpen
  \bibfield  {author} {\bibinfo {author} {\bibfnamefont {R.~J.}\ \bibnamefont
  {Francis-Jones}}\ and\ \bibinfo {author} {\bibfnamefont {P.~J.}\ \bibnamefont
  {Mosley}},\ }\href@noop {} {\bibfield  {journal} {\bibinfo  {journal} {arXiv
  preprint arXiv:1409.1394}\ } (\bibinfo {year} {2014})}\BibitemShut {NoStop}%
\bibitem [{\citenamefont {Bonneau}\ \emph {et~al.}(2015)\citenamefont
  {Bonneau}, \citenamefont {Mendoza}, \citenamefont {O'Brien},\ and\
  \citenamefont {Thompson}}]{bonneau2015effect}%
  \BibitemOpen
  \bibfield  {author} {\bibinfo {author} {\bibfnamefont {D.}~\bibnamefont
  {Bonneau}}, \bibinfo {author} {\bibfnamefont {G.~J.}\ \bibnamefont
  {Mendoza}}, \bibinfo {author} {\bibfnamefont {J.~L.}\ \bibnamefont
  {O'Brien}}, \ and\ \bibinfo {author} {\bibfnamefont {M.~G.}\ \bibnamefont
  {Thompson}},\ }\href@noop {} {\bibfield  {journal} {\bibinfo  {journal} {New
  Journal of Physics}\ }\textbf {\bibinfo {volume} {17}},\ \bibinfo {pages}
  {043057} (\bibinfo {year} {2015})}\BibitemShut {NoStop}%
\bibitem [{\citenamefont {Weston}\ \emph {et~al.}(2016)\citenamefont {Weston},
  \citenamefont {Chrzanowski}, \citenamefont {Wollmann}, \citenamefont
  {Boston}, \citenamefont {Ho}, \citenamefont {Shalm}, \citenamefont {Verma},
  \citenamefont {Allman}, \citenamefont {Nam}, \citenamefont {Patel} \emph
  {et~al.}}]{weston2016efficient}%
  \BibitemOpen
  \bibfield  {author} {\bibinfo {author} {\bibfnamefont {M.~M.}\ \bibnamefont
  {Weston}}, \bibinfo {author} {\bibfnamefont {H.~M.}\ \bibnamefont
  {Chrzanowski}}, \bibinfo {author} {\bibfnamefont {S.}~\bibnamefont
  {Wollmann}}, \bibinfo {author} {\bibfnamefont {A.}~\bibnamefont {Boston}},
  \bibinfo {author} {\bibfnamefont {J.}~\bibnamefont {Ho}}, \bibinfo {author}
  {\bibfnamefont {L.~K.}\ \bibnamefont {Shalm}}, \bibinfo {author}
  {\bibfnamefont {V.~B.}\ \bibnamefont {Verma}}, \bibinfo {author}
  {\bibfnamefont {M.~S.}\ \bibnamefont {Allman}}, \bibinfo {author}
  {\bibfnamefont {S.~W.}\ \bibnamefont {Nam}}, \bibinfo {author} {\bibfnamefont
  {R.~B.}\ \bibnamefont {Patel}},  \emph {et~al.},\ }\href@noop {} {\bibfield
  {journal} {\bibinfo  {journal} {Optics Express}\ }\textbf {\bibinfo {volume}
  {24}},\ \bibinfo {pages} {10869} (\bibinfo {year} {2016})}\BibitemShut
  {NoStop}%
\bibitem [{\citenamefont {Halder}\ \emph {et~al.}(2009)\citenamefont {Halder},
  \citenamefont {Fulconis}, \citenamefont {Cemlyn}, \citenamefont {Clark},
  \citenamefont {Xiong}, \citenamefont {Wadsworth},\ and\ \citenamefont
  {Rarity}}]{halder2009nonclassical}%
  \BibitemOpen
  \bibfield  {author} {\bibinfo {author} {\bibfnamefont {M.}~\bibnamefont
  {Halder}}, \bibinfo {author} {\bibfnamefont {J.}~\bibnamefont {Fulconis}},
  \bibinfo {author} {\bibfnamefont {B.}~\bibnamefont {Cemlyn}}, \bibinfo
  {author} {\bibfnamefont {A.}~\bibnamefont {Clark}}, \bibinfo {author}
  {\bibfnamefont {C.}~\bibnamefont {Xiong}}, \bibinfo {author} {\bibfnamefont
  {W.~J.}\ \bibnamefont {Wadsworth}}, \ and\ \bibinfo {author} {\bibfnamefont
  {J.~G.}\ \bibnamefont {Rarity}},\ }\href@noop {} {\bibfield  {journal}
  {\bibinfo  {journal} {Optics Express}\ }\textbf {\bibinfo {volume} {17}},\
  \bibinfo {pages} {4670} (\bibinfo {year} {2009})}\BibitemShut {NoStop}%
\bibitem [{\citenamefont {Francis-Jones}\ \emph {et~al.}(2016)\citenamefont
  {Francis-Jones}, \citenamefont {Hoggarth},\ and\ \citenamefont
  {Mosley}}]{francis2016all}%
  \BibitemOpen
  \bibfield  {author} {\bibinfo {author} {\bibfnamefont {R.~J.}\ \bibnamefont
  {Francis-Jones}}, \bibinfo {author} {\bibfnamefont {R.~A.}\ \bibnamefont
  {Hoggarth}}, \ and\ \bibinfo {author} {\bibfnamefont {P.~J.}\ \bibnamefont
  {Mosley}},\ }\href@noop {} {\bibfield  {journal} {\bibinfo  {journal}
  {Optica}\ }\textbf {\bibinfo {volume} {3}},\ \bibinfo {pages} {1270}
  (\bibinfo {year} {2016})}\BibitemShut {NoStop}%
\bibitem [{\citenamefont {Fang}\ \emph {et~al.}(2013)\citenamefont {Fang},
  \citenamefont {Cohen}, \citenamefont {Moreno},\ and\ \citenamefont
  {Lorenz}}]{fang2013state}%
  \BibitemOpen
  \bibfield  {author} {\bibinfo {author} {\bibfnamefont {B.}~\bibnamefont
  {Fang}}, \bibinfo {author} {\bibfnamefont {O.}~\bibnamefont {Cohen}},
  \bibinfo {author} {\bibfnamefont {J.~B.}\ \bibnamefont {Moreno}}, \ and\
  \bibinfo {author} {\bibfnamefont {V.~O.}\ \bibnamefont {Lorenz}},\
  }\href@noop {} {\bibfield  {journal} {\bibinfo  {journal} {Optics express}\
  }\textbf {\bibinfo {volume} {21}},\ \bibinfo {pages} {2707} (\bibinfo {year}
  {2013})}\BibitemShut {NoStop}%
\end{thebibliography}

\newpage

\section*{Appendix A: Algorithms}

\subsection*{Annealing}
\begin{enumerate}
 \setlength\itemsep{0em}
\item Define target phase matching function $\phi (\Delta k)$ and initial domain configuration.
\item Group together all the adjacent domains in bigger blocks having the same orientation.
\item Define algorithm temperature T, temperature step $\Delta T$ and energy threshold $E_t$ (used for accepting a given configuration as the optimal one).
\item Compute the energy for the initial configuration and store it in the variable $E_{\text{min}}$.
\item Apply a random perturbation to the blocks' width of up to $1\%$ of their current width.
\item Calculate the new phase matching function.
\item Compute the corresponding energy $E$.
\item If $E < E_{\text{min}}$, accept the current block configuration as the best configuration, update $E_{\text{min}} = E$ and go to step 11.
\item Else if $E>E_{\text{min}}$, accept the new configuration with probability $\exp (-E/T)$.
\item If the new configuration is not accepted, return to the best configuration an decrement $T$ by $\Delta T$. 
\item Iterate 5 to 10 until $E_{\text{min}}<E_t$ or $T=0$.
\end{enumerate}

\begin{figure*}[htb]\center
\includegraphics[width=1\textwidth]{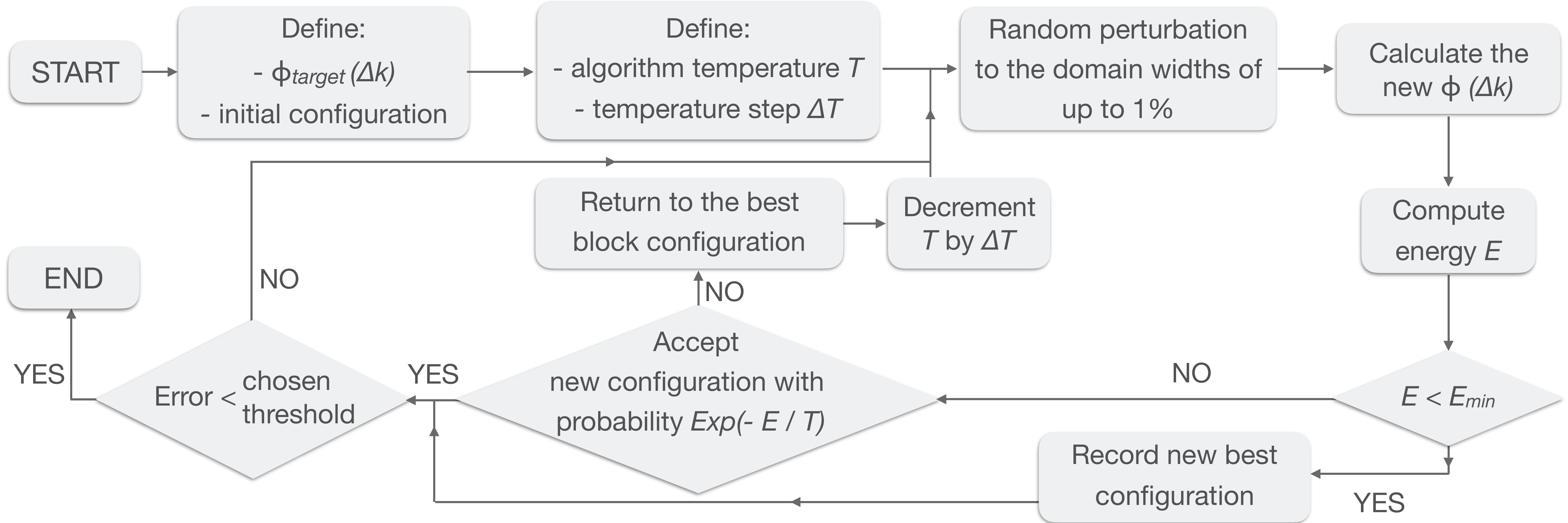}
\caption{
Block diagram of the simulated annealing algorithm.}
\label{AnnealingAlgorithm}
\end{figure*}

A block diagram of the algorithm is shown in fig. \ref{AnnealingAlgorithm}.

\subsection*{Arbitrary-small domains method}
\begin{enumerate}
 \setlength\itemsep{0em}
\item Define complex target amplitude profile.
\item Define parameters $w$, $\Lambda$, the number of domains $N$.
\item Initialize an empty list $S=\{\}$ for storing the domain orientation.
\item Define $m=1$.
\item Create two trial lists that are identical except for the last element: $S_\textsc{up} = S + \{\textsc{up}\}$ and $S_\textsc{down} =  S + \{\textsc{down}\}$
\item Compute cost functions for the two trial lists: $e_{\textsc{up}}=e_m(S_{\textsc{up}})$ and $e_{\textsc{down}}=e_m(S_{\textsc{down}})$.
\item If $e_{\textsc{up}}<e_{\textsc{down}}$, update $S=S_{\textsc{up}}$.
\item Else if $e_{\textsc{up}}>e_{\textsc{down}}$, update $S=S_{\textsc{down}}$.
\item Update $m=m+1$.
\item Iterate Steps 4 to 9 until $m=N+1$.
\end{enumerate}

\end{document}